\documentclass[twocolumn]{aastex631}

\usepackage{natbib}
\usepackage{graphicx}
\usepackage{hyperref}
\usepackage{amsmath}
\usepackage{lipsum}
\usepackage{grffile}
\usepackage{hyphenat}
\usepackage{xcolor}

\begin{document}

\title{The Absolute Age of NGC 3201 derived from Detached Eclipsing Binaries and the Hess Diagram}

\author{Jiaqi (Martin) Ying}
\affiliation{Department of Physics and Astronomy, Dartmouth College, 6127 Wilder Laboratory, Hanover, NH 03755, USA}

\author{Brian Chaboyer} 
\affiliation{Department of Physics and Astronomy, Dartmouth College, 6127 Wilder Laboratory, Hanover, NH 03755, USA}

\author{Wenxin Du} 
\affiliation{Department of Statistics, The Ohio State University, OH 43210, USA}

\begin{abstract}
    We estimate the absolute age of the globular cluster NGC 3201 using $10,000$ sets of theoretical isochrones constructed through Monte Carlo simulation using the Dartmouth Stellar Evolution Program.  These isochrones take into consideration of uncertainty introduced by the choice of stellar evolution parameters. We fit isochrones with 3 detached eclipsing binaries and obtained an age independent of  distance. We also fit isochrones with differential reddening corrected HST photometry data utilizing two different Hess diagram based fitting methods. Results from 3 different methods analyzing 2 different types of data agree to within $1 \sigma$, and we find the absolute age of NGC 3201 $= 11.85 \pm 0.74$ Gyr. We also perform variable importance analysis to study the uncertainty contribution from individual parameters and we find the distance is the dominance source of uncertainty in photometry based analysis while total metallicity, Helium abundance, $\alpha$-element abundance, mixing length, and treatment of helium diffusion are important source of uncertainties for all 3 methods.
\end{abstract}

\section{Introduction}
Globular clusters (GCs) are gravitationally bound clusters of stars. Even though the formation of GCs is still under debate\citep{forbes_globular_2018}, they host some of the oldest stellar populations in our Galaxy and can be used to constrain the age of the universe \citep{krauss_age_2003}.  Using JWST data, \citet{mowla_sparkler_2022} found proto-GCs formed at $z > 9$($\sim 0.5$ Gyr after the big bang) and \citet{adamo_discovery_2024} found young massive star clusters at $z \sim 10.2$ ($\sim 0.46$ Gyr after the big bang). Therefore, most GCs are relics of high-redshift star formation, and contain the fossil imprint of earliest phases of galaxy formation. As a result, they were widely used to probe the formation and assembly of the galaxy \cite[e.g.][]{kruijssen_kraken_2020}. Moreover, the numerous and ancient metal-poor GCs suggests that they might be the important contributors to ionizing radiation in the reionization era \citep{boylan-kolchin_little_2018}.

The Milky Way hosts $> 150$ GCs and there has been a substantial amount of studies using the observational data from HST \cite[e.g.][]{sarajedini_acs_2007,piotto_hubble_2015} and JWST \citep{ziliotto_multiple_2023}. 
To first order, stars in a GC can be assumed to form at same time with the same composition; as a result, theoretical isochrone age fitting is the most widely-used method to determine the age of GCs \citep[e.g][]{ying_absolute_2023, dotter_acs_2010, omalley_absolute_2017}. Theoretical isochrones can be generated by finding the common phase of stellar evolution shared by stellar evolution model with different mass \citep{dotter_dartmouth_2008}.

NGC 3201 is an ideal target for our study. It is a low galactic latitude GC about $4.55 \pm 0.20$ kpc from us \citep{vasiliev_gaia_2021}. It is also a metal poor GC with metallicity [Fe/H] $= -1.48 \pm 0.02$ \citep{magurno_chemical_2018}. Studies has shown that it is likely not to be form in-situ and was accreted as part of other galaxies \citep{belokurov_-situ_2024}. It is a well studied GC mostly due to its richness in variable stars \cite[e.g.][]{layden_photometry_2003, kaluzny_clusters_2016, cortes_variability_2023}. Recently, \citet{giesers_detached_2018} found a detached stellar-mass black hole candidate in NGC 3201 using the radial velocity measurements of stars about unseen companions. \citet{rodriguez_constraints_2023} suggests that the mass function of the two (or potentially three) black holes in NGC 3201 can be used to place strong constraints on the cosmological coupling between black holes and an expanding universe. The accuracy of such models relies the estimated absolute age of NGC 3201 which is the goal of this study. 

\citet{ying_absolute_2023} has demonstrated that the absolute age of GCs can be determined by combining the deep HST Advanced Camera for Surveys (ACS) data \citep{sarajedini_acs_2007,anderson_acs_2008} and state-of-art Dartmouth Stellar Evolution Program \citep{dotter_dartmouth_2008} with Monte Carlo input stellar parameter (without assuming a fixed distance and reddening) through a number density based 2D CMD-fitting method. \citet{ying_absolute_2023} performed an careful analysis of the source of uncertainties in the estimate of the absolute age of GC M92, and found that the distance is the dominant source of uncertainty. Fortunately, NGC 3201 also hosts detached eclipsing binaries (DEBs). Because of its eclipsing nature, \citet{rozyczka_cluster_2022} was able to determine the radius, mass and luminosity of those DEBs without any prior information about the distance. We can model DEBs as single stars using our stellar evolution models, and compare those parameters with observational data without any assumptions of distance.

In this paper, we fit two sets of independent observational datasets of NGC 3201: HST ACS photometry and 3 DEBs with $10,000$ sets of theoretical isochrones through Monte Carlo simulation using the Dartmouth Stellar Evolution Program to measure the absolute age of NGC 3201. In \S \ref{Observational data}, we introduce the observational data;  \S \ref{Isochrone construction} covers the process of isochrone construction; \S \ref{Detached Eclipsing Binaries Isochrone Fitting} presents the details of our isochrone age fitting method utilizing Detached Eclipsing Binaries; \S \ref{Color Magnitude Diagram Isochrone Fitting} presents the details of two isochrone age fitting methods for utilizing the Color-Magnitude Diagrams with the Voronoi-Binning methods focus on the overall change in number densities and 2D Kolmogorov–Smirnov method focus on the largest discrepancy in the morphology; and \S \ref{Results} presents our main results and compare the result from different methods utilizing different observational data.

\section{Observational data} \label{Observational data}
\subsection{Detach Eclipsing Binaries}

\begin{table}[]
\caption{Detached Eclipsing Binaries from \citet{rozyczka_cluster_2022} \label{tab2}}
\begin{tabular}{llll}
\hline
& Mass  & Luminosity    & Radius                                  \\
&($M_{\odot}$) &($L_{\odot}$) &($R_{\odot}$)                                  \\
\hline
V138p & $0.784^{+5}_{-5}$ & $1.40^{+28}_{-36}$ & $0.973^{+45}_{-45}$ \\
V138s & $0.716^{+3}_{-3}$ & $0.61^{+20}_{-20}$ & $0.760^{+50}_{-71}$ \\
V139p & $0.806^{+7}_{-7}$ & $1.96^{+24}_{-24}$ & $1.215^{+5}_{-12}$ \\
V139s & $0.684^{+3}_{-3}$ & $0.40^{+6}_{-6}$ & $0.687^{+4}_{-4}$ \\
V141p & $0.838^{+7}_{-8}$ & $4.22^{+49}_{-56}$ & $2.458^{+55}_{-103}$ \\
V141s & $0.724^{+5}_{-5}$ & $0.57^{+6}_{-14}$ & $0.750^{+21}_{-67}$ \\
\hline
\end{tabular}
\end{table}

\citet{rozyczka_cluster_2022} provides a careful analysis of four sets of DEBs in NGC 3201, 
and suggest that one DEB: V142 seems to evolve along a different path than the remaining three (likely with a non-standard history and/or chemical composition). Therefore, we utilize only 3 DEBs in our analysis with their Mass, luminosity and radius shown in Table.\ref{tab2}. 

\subsection{Photometric Data}
To estimate the age of NGC 3201, we use calibrated photometric data for NGC 3201 from the Hubble Space Telescope (HST) Advanced Camera for Surveys (ACS) globular cluster survey treasury program \citep{sarajedini_acs_2007,anderson_acs_2008}.
The ACS GC survey included artificial star tests that provide an estimate of the photometric uncertainties and completeness as a function of magnitude and cluster position \citep{anderson_acs_2008}. 
We use a subset of stars around the main sequence turn-off (MSTO) to fit isochrones whose position is most sensitive to variations in age, and relatively insensitive to the present day mass function \citep{chaboyer_accurate_1996}.  These stars have a $15.45 < \mathrm{F606W} < \mathrm{19.45}$, which is $\pm 2\,$magnitudes of the point on the subgiant branch which is $0.05\,$mag redder than MSTO. Additionally, we remove blue straggler stars and outliers by selecting stars that are within $0.08\,$mag in F606W of the median ridgeline in an magnitude-magnitude diagram of F814W and F606W. With these cuts, our observational sample contains  $4,686$ stars.

NGC 3201 is not a low reddening cluster as the average reddening for NGC 3201 is  $E (B - V) = 0.24$ \citep{harris_catalog_1996}. \citet{legnardi_differential_2023} studied the effect of differential reddening on the CMD of GCs and shows that NGC 3201 exhibits a high differential reddening: $\sigma_{\Delta \mathrm{F606W}} = 0.022 \pm 0.002$. Differential reddening introduces significant broadening effect in CMD which can compromise the precision of our number density based CMD-fitting method \citep{ying_absolute_2023}. 


\begin{figure*}
    \centering
    \includegraphics[width=\textwidth]{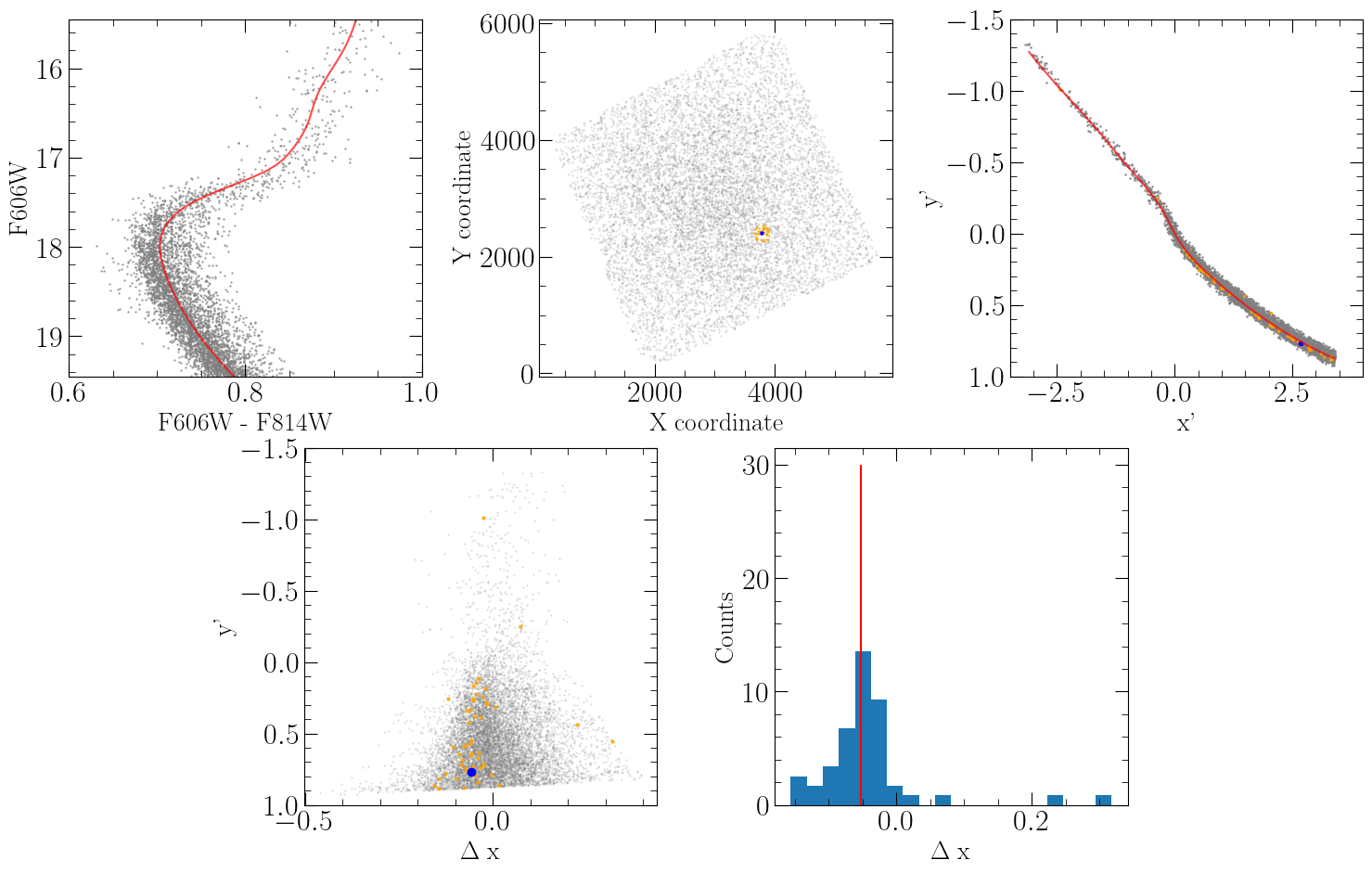}
    \caption{Illustration of differential reddening correction process. Top left (a): The fiducial ridgeline for NGC 3201(in red) generated from CMD(in grey) using the fidanka package. Top Middle (b): Position of NGC 3201 stars in the ACS/WFC field of view (in grey). The selected target star (in blue) and its 50 neighbors (in orange).
    Top Right (c): NGC 3201 stars on the rotated CMD (in grey) with rotated fiducial ridgeline (in red), target star (in blue) and its 50 neighbors (in orange).
    Bottom left (d): $\Delta x'$ or change in abscissa for NGC 3201 stars (in grey), target star (in blue) and its 50 neighbors (in orange). 
    Bottom Right (e): Distribution of $\Delta x'$  for the target star (blue histogram) and the median value (in red).}
    \label{fig:DRCR}
\end{figure*}

Thus, a differential reddening correction for the photometric data is performed using the following procedure (inspired by \citet{milone_acs_2012}:
\begin{enumerate}
    \item Extract the fiducial ridgeline from the CMD using the fidanka package \citep{boudreaux_fidanka_2023} 
    as shown in Figure \ref{fig:DRCR}a 
    \item A new reference system is defined with $x$-axis: abscissa being parallel to the reddening line. To achieve that, we rotate the CMD counterclockwise by angle:
    \begin{equation*}
        \theta = \frac{A_{\mathrm{F606W}}}{A_{\mathrm{F606W}} - A_{\mathrm{F814W}}},
    \end{equation*}
    $A_{\mathrm{F606W}}$ and $A_{\mathrm{F814W}}$ are the absorption coefficients in the F606W and F814W ACS bands. We adopt $E(B - V) = 0.24$ \citep{harris_catalog_1996} as the average reddening and assume a cool star with $T_{ref} = 4000 K$. \cite{bedin_transforming_2005} provides the absorption coefficient for NGC 3201 as $A_{\mathrm{F606W}} = 0.588$ and $A_{\mathrm{F814W}} = 0.441$. The rotation angle is: $\theta = 1.249$ rad.
    \item For each star on the rotated CMD, the $50$ stars with shortest spatial distance were selected to be the target stars neighbors as shown in Figure \ref{fig:DRCR}b. 
    \item For each of its neighbors, the difference in abscissa between it and the rotated fiducial ridgeline (shown in Figure \ref{fig:DRCR}c) is calculated. Figure \ref{fig:DRCR}d shows a significant bias towards negative value in difference in abscissa. This is expected as the differential reddening is highly correlated with the spatial location of the star. Figure \ref{fig:DRCR}e shows the distribution of the difference in abscissa. We assign the median $\Delta x'$ as the differential reddening of the target star to avoid the influence of binaries.
    \item The abscissa of each target stars is subtracted by the median $\Delta x'$ of its neighbors to correct for differential reddening. The resulted CMD is rotated clockwise by $\theta$ to restore the original coordinate system.
\end{enumerate}
\begin{figure}
    \centering
    \includegraphics[width=0.45\textwidth]{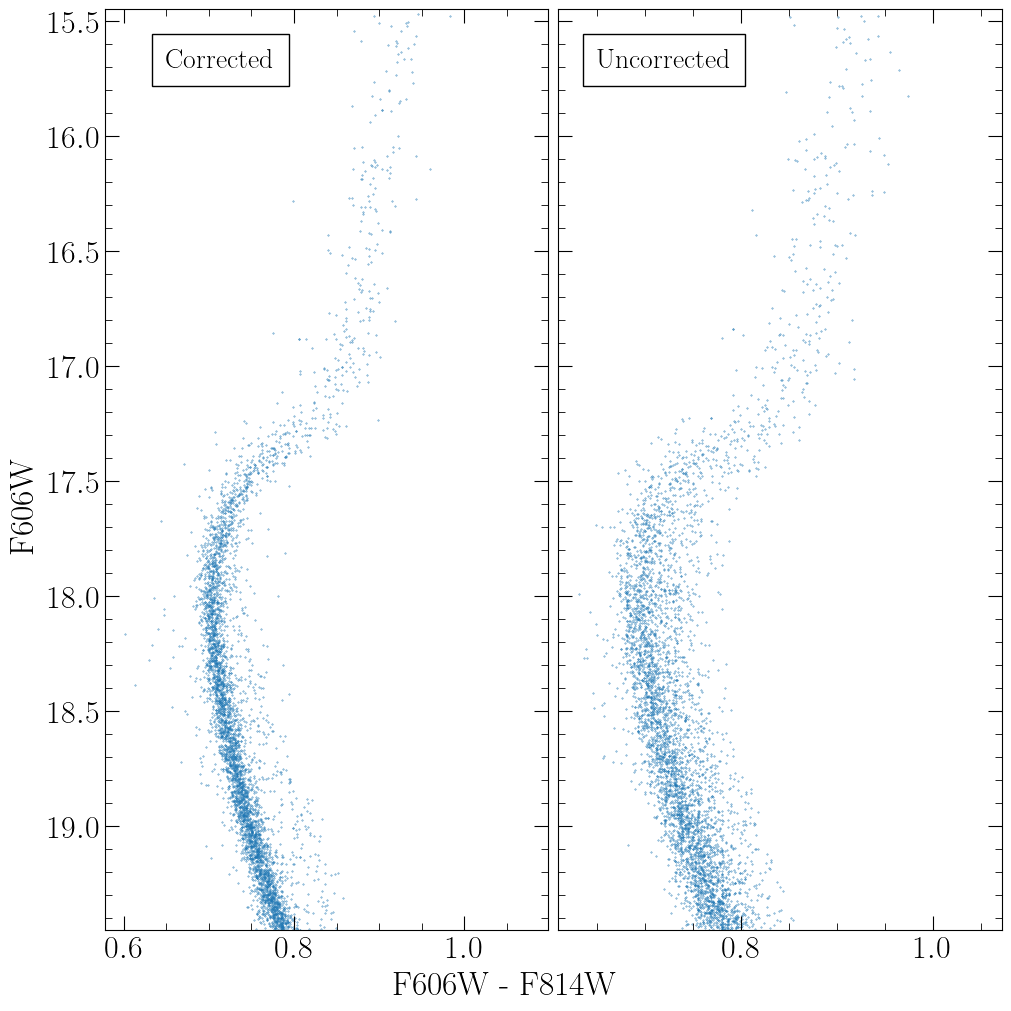}
    \caption{Comparison of CMDs. Left: CMD after correction for differential reddening. Right: CMD before correction for differential reddening. }
    \label{fig:DRCR_cmd}
\end{figure}
Figure \ref{fig:DRCR_cmd} shows the result of differential reddening correction for NGC 3201. The broadening effect caused by differential reddening is significantly reduced. We use the differential reddening corrected CMD for NGC 3201 as the observation data in this study.

We note that NGC 3201, like other old globular clusters, hosts multiple stellar populations \cite{carretta_na-o_2009}. Because multiple stellar populations typically presents abundance variations in light elements such as C, N, O and etc., \citet{milone_hubble_2017} uses HST UV Globular Cluster Survey \citep{piotto_hubble_2015} and finds 2 populations of NGC 3201. However, \citet{vandenberg_models_2022} suggests this phenomenon is hardly observable in red filters such as the F606W and F814W data used in this paper as isochrones looks almost identical with enhanced light elements. More importantly, multiple populations in globular clusters have very insignificant age difference \citep[e.g.]{lucertini_hubble_2021, ziliotto_multiple_2023}. As a result, these multiple populations will not be considered in this study.

\section{Isochrone construction} \label{Isochrone construction}

\begin{table*}[]
\caption{Monte Carlo Input Parameters \label{tab1}}
\begin{tabular}{llll}
\hline
Variable                               & Distribution & Range        & Source                                               \\
\hline
{[}Fe/H{]}                    & Normal       & $-1.48 \pm 0.07$&  \cite{magurno_chemical_2018}\\
 & & & \cite{carretta_intrinsic_2009} \\
 & & & \cite{harris_catalog_1996} \\
{[}$\alpha$/Fe{]}             & Normal       & $0.37 \pm 0.07$& \cite{magurno_chemical_2018}                                               \\
 & & & \cite{rozyczka_cluster_2022} \\
$\Delta Y/\Delta Z$ & Uniform & $1.75 \sim 2.5$ & \citet{peimbert_primordial_2016}\\
Helium abundance   & Uniform      & $0.2465(25) + \left(\Delta Y/\Delta Z\right) Z$   & \citet{aver_effects_2015}                                       \\
Mixing length                 & Uniform      & $1.0 \sim 2.5$      & N/A                                         \\
Heavy element diffusion       & Uniform      & $0.5 \sim 1.3$      & \citet{thoul_element_1994}                                         \\
Helium diffusion              & Uniform      & $0.5 \sim 1.3$    & \citet{thoul_element_1994}                                           \\
Surface boundary condition 
 &   Trinary          &     1/3; 1/3; 1/3             & \citet{eddington_internal_1926}\\
  &             &                     & \citet{krishna_swamy_profiles_1966} \\
  &             &                     & \citet{hauschildt_nextgen_1999} \\
Low temperature opacities     & Uniform      & $0.7 \sim 1.3$        & \citet{ferguson_low-temperature_2005}                                      \\
High temperature opacities    & Normal       & $1.0 \pm 0.03$           & \citet{iglesias_updated_1996}     \\
Plasma neutrino loses         & Normal       & $1.0 \pm 0.05$     & \citet{haft_standard_1994}     \\
Conductive opacities          & Normal       & $1.0 \pm 0.20$       & \citet{hubbard_thermal_1969}\\
 &             &                     & \citet{canuto_electrical_1970}   \\
Convective envelope overshoot & Uniform      & $0 \sim 0.2$      & N/A                                          \\
Convective core overshoot & Uniform & $0 \sim 0.2$ & N/A \\
$p + p \to H_2 + e + \nu$      
& Normal       & $\left(4.07 \pm 0.04 \right)\times 10^{-22}$ & \citet{acharya_uncertainty_2016}\\
 & & & \citet{marcucci_proton-proton_2013}\\
${ }^{3}He + { }^{3}He \to { }^{4}He + p + p$                 & Normal       & $5150 \pm 500$& \citet{adelberger_solar_2011}\\
${ }^{3}He + { }^{4}He \to { }^{2}H + \gamma$                  & Normal       & $0.54 \pm 0.03$&\citet{deboer_monte_2014}\\
${ }^{12}C + p \to { }^{13}N + \gamma$                & Normal       & $1.45 \pm 0.50$ & \citet{xu_nacre_2013}\\
${ }^{13}C + p \to { }^{14}N + \gamma$              & Normal       & $5.50 \pm 1.20$& \citet{chakraborty_systematic_2015}\\
${ }^{14}N + p \to { }^{15}O + \gamma$             & Normal       & $3.32 \pm 0.11$ & \citet{marta_n14pensuremathgammao15_2011}\\
${ }^{16}N + p \to { }^{17}F + \gamma$               & Normal       & $9.40 \pm 0.80$ & \citet{adelberger_solar_2011}\\
\hline
\end{tabular}
\end{table*}

We use the Dartmouth Stellar Evolution Program (DSEP) \citep{dotter_dartmouth_2008} to generate stellar models and generally use literature estimates when adopting uncertainties for each parameter (see Table~\ref{tab1}, and discussion in \citet{ying_absolute_2023}). We adopt theoretical or experimental uncertainties for most of the variables listed in Table~\ref{tab1}. There are, however, some variables with uncertainties which are not easily quantifiable. For example, the mixing length theory of convection remains the dominant framework for one-dimensional (1D) convective energy transport used in stellar structure and evolution calculations \citep{joyce_review_2023} and nearly all models use a solar-calibrated mixing length even though a wide range of studies have shown that it is inappropriate to adopt the solar-calibrated mixing length ad hoc in any given stellar model \citep[e.g.][]{guenther__2000, joyce_not_2018}. Unfortunately, there is no clear relationship between mixing length and other variables such as mass, metallicity and etc.. As a result, \citet{ying_absolute_2023} adopt a wider input range for the mixing length parameter $\alpha_{\textup{MLT}}$ to cover the range of empirical calibrated mixing length parameter $\alpha_{\textup{MLT}}$ in stellar evolution models and shows the mixing length is one of the major source of uncertainty in stellar evolution models and is highly correlated with models of atmosphere. We adopt the same range of the mixing length parameter $\alpha_{\textup{MLT}}$ for this study.

We generate $10,000$ sets of input parameters by doing Monte Carlo simulations on parameters shown in Table~\ref{tab1} from their associated probability distribution functions. Each set of input parameters is used to evolve 13 low-mass stellar models with mass from $0.2\,M_{\odot}$ to $0.68\,M_{\odot}$ with an increment of $0.04 \,M_{\odot}$, 14 medium-low-mass stellar models with mass from $0.7\, M_{\odot}$ to $1.35 \,M_{\odot}$ with an increment of $0.05 \,M_{\odot}$, 6 medium-high-mass stellar models with mass from $1.4\, M_{\odot}$ to $1.9 \,M_{\odot}$ with an increment of $0.1 \,M_{\odot}$, and 6 high-mass stellar models with mass from $2.0\, M_{\odot}$ to $3.0 \,M_{\odot}$ with an increment of $0.2 \,M_{\odot}$. The lower-mass models use the FreeEOS-2.2.1 \citep{irwin_freeeos_2012}, while the higher-mass models use an analytical equation of state which include the Debyre-Huckel correction \citep{chaboyer_opal_1995}. \citet{dotter_mesa_2016} describes a robust method to transform a set of stellar evolution tracks onto a uniform basis and then interpolate within that basis to construct stellar isochrones. We adopt this equivalent evolutionary phase(EEP)-based method to generate $41$ theoretical isochrones from $8\,$Gyr to $16\,$Gyr with an increment of $200\,$Myr. 
Each isochrone is  constructed with a dense grid of $400$ EEPs in order to ensure that the output isochrones have a high density of points to avoid any interpolation errors when constructing simulated color-magnitude diagrams (sCMDs). In summary, we generated $10,000$ isochrone sets.  Each isochrone set consists of  $41$ isochrones of different ages, for a total of $10,000 \times 41 = 410,000$ individual isochrones. Those isochrones are used to fit DEBs directly while the Simulated Color-Magnitude Diagram generated based on those isochrones are used to fit photometric data.

\subsection{Simulated Color-Magnitude Diagram}
Each MC set of theoretical isochrones are used to create a set of simulated color-magnitude diagrams (sCMDs) of NGC 3201 which will be used to compare with the observational CMD of  \cite{sarajedini_acs_2007}. A sCMD is constructed by randomly creating a $4,000,000$ samples with present-date mass function $=-1.22$ \citep{ebrahimi_new_2020} and binary fraction $=0.061$ \citep{milone_acs_2012} for each isochrone as described in detail by \citet{ying_absolute_2023}.  In brief this procedure combines the theoretical isochrones with the observed GC density profile, present day mass function, binary mass fraction, photometric completeness and photometric errors to generate a sCMD that accurately reflects the properties of the observed CMD. 


\section{Detached Eclipsing Binaries Age Fitting Method} \label{Detached Eclipsing Binaries Isochrone Fitting}
\subsection{Isochrone Fitting}
The ages of the three DEBs were determined by comparing the observed 
mass, luminosity and radius  \citep{rozyczka_cluster_2022} to the predicted values in the theoretical isochrones.
Table~\ref{tab2} lists observational information for all 6 stars studies in \citet{rozyczka_cluster_2022}. 
We define the following metric to determine goodness-of-fit between a point on the isochrone and an observed star:
\[
\chi^2_{i,j,k} = \sum_{i} \frac{\left(o_{i,j} - t_{i,k} \right)^2}{\sigma_{o_{i,j}}^2},
\]
where $o_{i,j}$ are the values for the $i$-th parameter of the $j$-th star derived from observations, $t_{i,k}$ are the values assumed for the $i$-th parameter in the $k$-th point in models (theory), and $\sigma_{o_{i,j}}$ are the observational uncertainties for the $i$-th parameter of the $j$-th star. In this case, we fit three parameters simultaneously. The goodness-of-fit of $k$-th point in isochrone for $j$-th star is (where the subscripts $M$,$R$ and $L$ represent mass, luminosity and radius):
\[
\chi^2_{j,k} = \frac{\left(o_{M,j} - t_{M,k}\right)^2}{\sigma_{o_{M,j}}^2} + \frac{\left(o_{L,j} - t_{L,k}\right)^2}{\sigma_{o_{L,j}}^2} + \frac{\left(o_{R,j} - t_{R,k}\right)^2}{\sigma_{o_{R,j}}^2},
\]
for each star. 
For each isochrone, we find:
\begin{equation}
\chi^2_{\textup{iso}} = \sum_{\textup{star} j} \min_{k \in \textup{iso}}\left\{ \chi^2_{j,k} \right\} \label{eq:chi2_iso},
\end{equation}
which is the sum of minimal $\chi^2$ value for each star and assign $\chi^2_{\textup{iso}}$ as the goodness-of-fit for the isochrone. A similar metric have been used in a variety of studies and has been proven effective \citep[e.g.][]{omalley_absolute_2017, joyce_ages_2023}. 

\begin{figure}
    \centering
    \includegraphics[width=0.45\textwidth]{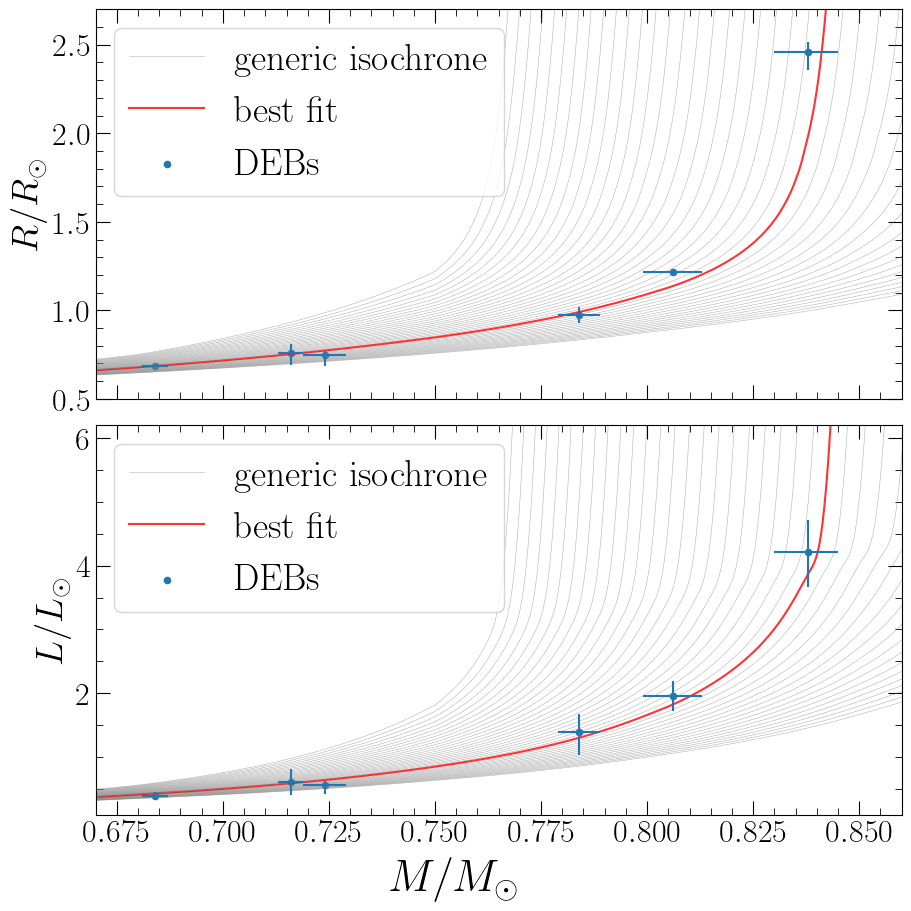}
    \caption{Comparing a set of theoretical isochrones with different age from $8$ Gyr to $16$ Gyr on the Mass-Luminosity-Radius space. Blue points are the observational data with corresponding uncertainties. The best fit isochrone (in red) which has age $=11.4$ Gyr has a lower $\chi^2_{\textup{iso}}$ value (calculated using equation (\ref{eq:chi2_iso})) comparing to other isochrones (in grey). Top: isochrones and DEBs on mass vs radius plane. Bottom: isochrones and DEBs on mass vs luminosity plane.}
    \label{fig:rml_demo}
\end{figure}

Figure \ref{fig:rml_demo} shows an example of fitting DEBs with theoretical isochrones in the Mass-Luminosity-Radius space. For each isochrone, we calculate its $\chi^2_{\textup{iso}}$ using equation (\ref{eq:chi2_iso}) as its goodness-of-fit.

\subsection{Bootstrap resampling} \label{DEBs bootstrap}
We notice that even though we define our testing metric $\chi^2_{\textup{iso}}$ similarly to the $\chi^2$ goodness-of-fit, $\chi^2_{\textup{iso}}$ does not necessarily follow the $\chi^2$ distribution. A $\chi^2$ distribution is defined by the degree-of-freedom $k$ which is the number of independent variables. In this case, however, $k$ is not well defined. We can treat six observed stars as independent but the three parameters: mass, luminosity and radius of each star are not independent. In a theoretical isochrone, as shown in Figure~\ref{fig:rml_demo}, there is clear correlation between mass, luminosity and radius. As a result, ideally, $\chi^2_{\textup{iso}}$ should be modeled as the sum of squares of dependent Gaussian random variables and the distribution can be estimated with its covariance matrix. Since the covariance matrix of the observational data is not provided, we cannot estimate a closed form distribution for $\chi^2_{\textup{iso}}$. 

An alternative solution is to empirically estimate the distribution of $\chi^2_{\textup{iso}}$ through bootstrapping in the following steps:
\begin{enumerate}
    \item Pick an ``good-fit'' isochrone with a low $\chi^2_{\textup{iso}}$ as underlying population.
    \item Simulate 6 stars by randomly sampling from the isochrone and add observational uncertainty corresponding to that star listed on Table~\ref{tab2}.
    \item Calculate $\chi^2_{\textup{iso,sim}}$ using equation (\ref{eq:chi2_iso}).
    \item Repeat the process for $10,000$ times.
\end{enumerate}

\begin{figure}
    \centering
    \includegraphics[width=0.45\textwidth]{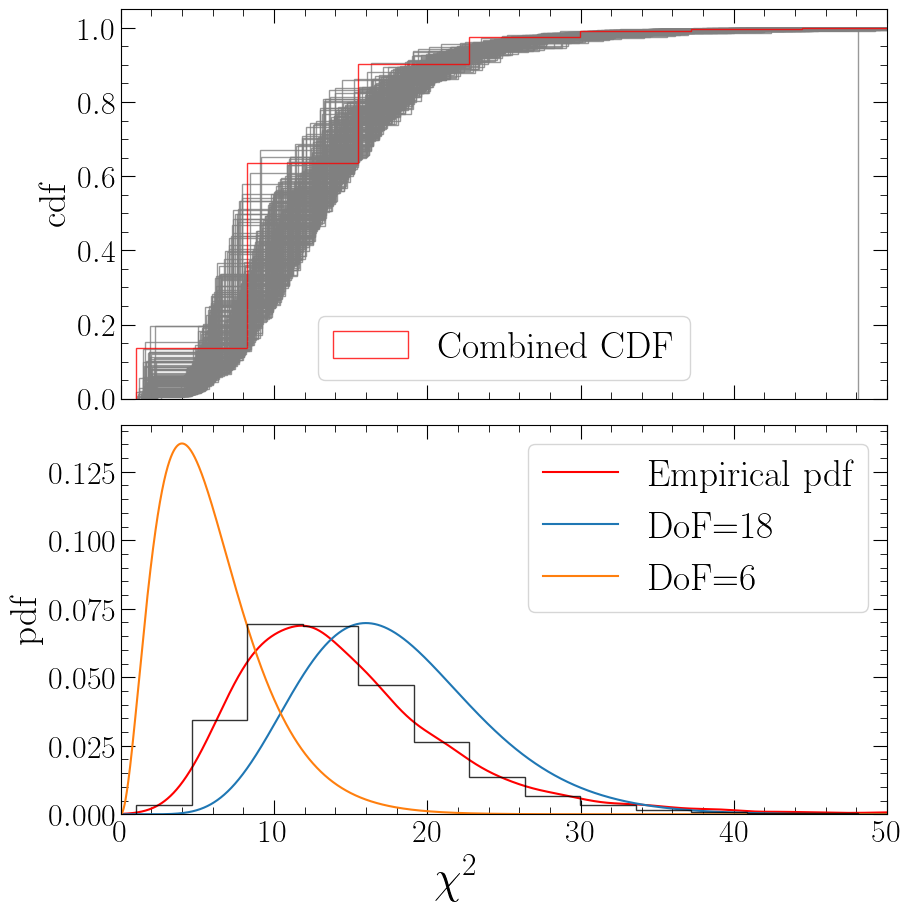}
    \caption{Bootstrap resampling for $\chi^2_{\textup{iso}}$. Top: cumulative distribution for $\chi^2_{\textup{iso,sim}}$ of $200$ different isochrones (in grey) and the combined cumulative distribution (in red). Bottom: probability distribution for $\chi^2$ distribution with degree-of-freedom $=18$ (in blue), and degree-of-freedom $=6$ (in orange) comparing with the estimated empirical probability distribution for $\chi^2_{\textup{iso}}$ (in red).}
    \label{fig:rml_resample}
\end{figure}

Figure~\ref{fig:rml_resample} shows the result of bootstrap resampling. We first investigate the choice of ``good-fit'' on the estimated distribution for $\chi^2_{\textup{iso}}$. We select $200$ isochrones with different input parameters and age and plot the distribution for each of them on the top panel of Figure~\ref{fig:rml_resample}. We observe that they follow the same trend and the individual difference is likely to be caused by the randomness in resampling. We combine the result from those $200$ isochrone and use it as the empirical distribution for $\chi^2_{\textup{iso}}$. We compare the empirical distribution with $\chi^2$ distribution with degree-of-freedom $k=6$ and degree-of-freedom $k=18$. As we expected, empirical distribution is in between of those two distributions suggesting that the three parameters from a star are neither perfectly correlated ($k=6$) nor independent ($k=18$).

\section{Hess Diagram Isochrone Fitting} \label{Color Magnitude Diagram Isochrone Fitting}
\subsection{Voronoi-Binning Method} \label{Voronoi-Binning Method}
\citet{ying_absolute_2023} presented a new isochrone fitting method which fits the density of points within the CMD (ie.\ the fit is made to a Hess diagram). The method partitions the sCMD using an adaptive Voronoi-Binning technique  \citep{cappellari_adaptive_2003} which keeps the number of stars within a bin to be roughly constant.  The expected number of stars in a given bin is then compared to the observed number of stars.  A parametric bootstrapping method is used to resample the observed data using the photometric error and completeness from the artificial star test\citep{anderson_acs_2008} to generate an empirical  $\chi^2$ distribution which is used to determine the goodness-of-fit of a given isochrone to the observed data. 

\citet{ying_absolute_2023} estimated the absolute age of M92 $=13.80 \pm 0.75$ Gyr using this method, which is twice as  accurate as  the age estimated in \citet{omalley_absolute_2017} with a similar Monte-Carlo isochrone constructing method but with only using the MSTO as an age indicator.  This demonstrates the capability of a full Hess diagram fitting method.  Although very different in detail, we note that \citet{valcin_inferring_2020} also developed a method to determine the age of a GC by  fitting the Hess diagram.

Because the Voronoi-Binning method is a full CMD-fitting method, it is extremely sensitive to the morphological changes in anywhere on the CMD. As a result, it is extremely selective. \citet{ying_absolute_2023} showed that only $1,100$ isochrones out of the total $820,000$ isochrones generated fit the observational data.

We apply the same method to differential reddening corrected photometric data for NGC 3201. We collect literature values for estimated distance modulus using methods such as CMD fitting, RR Lyrae stars, DEBs, and etc., and estimated reddening using methods such as CMD fitting, RR Lyrae stars, dust maps, and etc. We summarize the result in Table~\ref{tab3} and adopt a wide range of distance modulus $(m - M)_{F606W} = 14.0 \sim 14.3$, and reddening $E(F606W - F814W) = 0.20 \sim 0.30$ as our prior and we use Gaussian process method to search for the combination of distance modulus and reddening with return the lowest $\chi^2$ value for each isochrone. NGC 3201 has a much lower stellar density compared to M92. As a result, only $4,664$ stars were selected for this analysis while \citet{ying_absolute_2023} utilized $18,077$ stars in M92. To compensate the decrease in number of observed stars, we reduce the number of Voronoi bins from $800$ in our M92 study to $200$, increasing the average number of stars in each bin. This, however, reduces the resolution of our Voronoi diagram.

\subsection{Bootstrap Resampling} \label{Vorbin bootstrap}
\citet{lin_research_2013} demonstrated that with large data sets, using the p-value-based hypothesis testing method no longer provides scientifically reliable results. Instead, we estimate the empirical $\chi^2$ distribution using bootstrap resampling as described in \citet{ying_absolute_2023}. Instead of fitting the observed data, we create ``fake'' observation data by sampling $4,664$ points from a sCMD generated with theoretical isochrones. By comparing the ``fake''  observation data with another sCMD generated from the same isochrone, the $\chi^2_{\textup{sim}}$ we determined show the intrinsic uncertainty caused by the randomness in photometric error in the situation where the model coincides with underlying population. 

\begin{figure}
    \centering
    \includegraphics[width=0.45\textwidth]{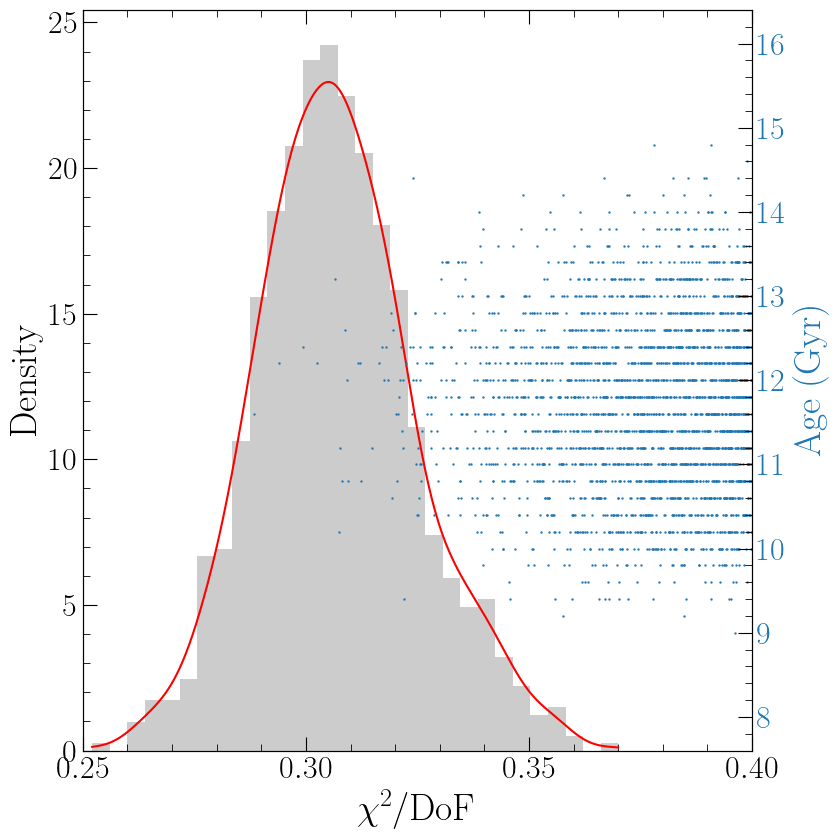}
    \caption{Bootstrap resampling for Voronoi-Binning method. The empirical $\chi^2$ distribution from $10,000$ bootstrap resampling is shown in the grey histogram with estimated kernel density function in red. Blue dots shows the distribution of $\chi^2$ when fitting theoretical isochrones with observed data.}
    \label{fig:vorbin_resample}
\end{figure}

The empirical $\chi^2$ distribution is determined by repeating the bootstrap resampling for $10,000$ times. Figure~\ref{fig:vorbin_resample} shows the distribution of empirical $\chi^2$, and is normalized by the number of stars in observational data which is the degree-of-freedom. We compare empirical $\chi^2$ distribution with 
$\chi^2$ from fitting theoretical isochrones with observed data to assign weight in the final analysis. We notice that only $2,321$ out of $410,000$ isochrones ($\approx 0.57\%$) has a non-zero weight and will be considered as a ``good-fit''.

\subsection{2D Kolmogorov–Smirnov method}
\citet{anderson_acs_2008} stated that there are two main sources of photometric error: the presence of other stars and errors in the modeled point spread function (PSF). The algorithm used to perform the photometry can be severally compromised by the presence of neighbors. As a result, the photometric uncertainties for GCs with high number density of stars such as M92 are dominated by crowding. This photometric uncertainty is captured by artificial star tests, which inject a small number of stars with a known magnitude into the real data to determine the photometric uncertainty and completeness.  

For GCs with a relatively low density of stars, imperfect PSF modeling can be the dominant source of photometric uncertainty.  However, the artificial star tests  use exactly the same PSF to inject the stars that is used measure the stars and so do not capture the uncertainty associated with the miss-match between the real and modeled PSF. It is impossible to quantify the uncertainty due to the PSF modeling without redesigning the artificial star test process\footnote{\citet{ying_absolute_2024} will provide more details and discussions about this problem.} which is beyond the scope of this paper.  The Voronoi-Binning method of age determations relies upon the assumption that the artificial star tests provide a reliable estimate of the photometric uncertaintiy in the observed data, which may not be the case in NGC 3201.  

As a result, we develop a fitting method which is more robust to the photometric uncertainty while maintaining the advantage of  Hess diagram fitting method. 2D Kolmogorov–Smirnov (KS) test is the multivariate extension of the famous Kolmogorov–Smirnov test, and a goodness-of-fit test which can be used to test for consistency between the empirical distribution of data points on a plane and a hypothetical density law \cite[e.g.][]{peacock_two-dimensional_1983, fasano_multidimensional_1987}. We develop a Hess diagram fitting method based on the 2D-KS test as a validation for the Voronoi-Binning Method described in Section~\ref{Voronoi-Binning Method}.

\begin{figure}
    \centering
    \includegraphics[width=0.45\textwidth]{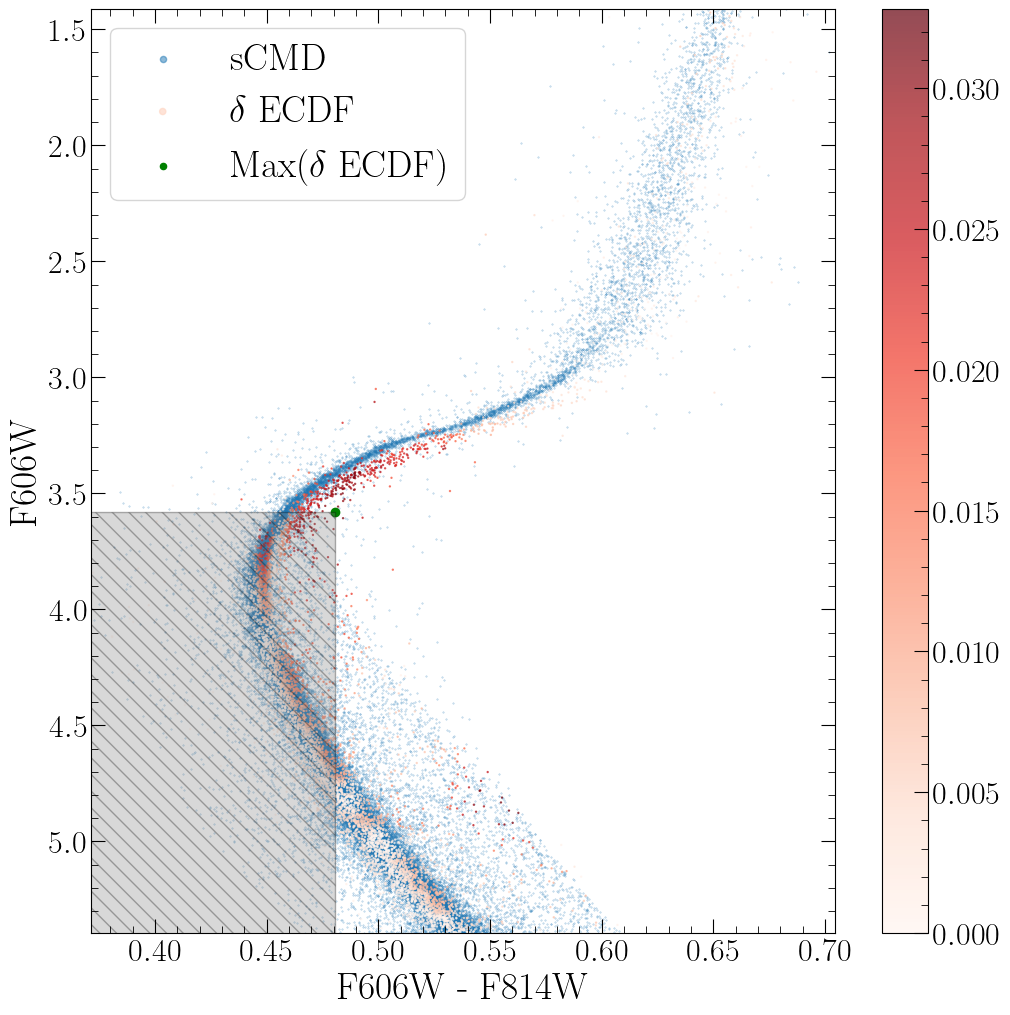}
    \caption{An example of 2D-KS method. sCMD is shown in blue dots. Green dot represent an observed stars from HST ACS data. The empirical cumulative distribution function(ECDF) at that point is calculated by dividing the number of observed stars within the grey shaded region by the total number of observed stars on the CMD. The red dot represent the difference in ECDF between the CMD and sCMD.}
    \label{fig:2dks}
\end{figure}

The process can be summarized as follows:
\begin{enumerate}
    \item Choose a set of distance modulus and reddening values and apply correction to the observed CMD to align with the sCMD.
    \item Estimate the empirical cumulative distribution function(ECDF) of the sCMD as the expected ECDF for the isochrone through using a divide and conquer algorithm \citep{bentley_multidimensional_1980}. Generate a linear interpolation of the expected ECDF to cover the CMD plane.
    \item Apply the same method to determine observed ECDF from the observed CMD.
    \item Compare observed ECDF and expected ECDF at the location of each observed stars and return the maximum difference.
    \item Predict the next set of distance modulus and reddening values base on Gaussian process within the predetermined boundary values and rerun the analysis process.
\end{enumerate}
Figure \ref{fig:2dks} shows an example of fitting an sCMD generated from an isochrone onto the observed CMD. For each star such as the green observed stars, the ECDF is calculated by counting the number of stars within the grey shaded region divided by the total number of stars. The red dots show the difference between expected ECDF and observed ECDF. In most case such as Figure \ref{fig:2dks}, the difference in ECDFs is highest at MSTO region where is most sensitive to change in age \cite{krauss_age_2003}. In this case, maximum difference $\approx 3\%$ which is the combined effect of morphological change at MSTO and inaccurate photometric binary fraction estimation.

The 2D-KS method addresses several problem with the Voronoi-Binning method mentioned before. For example, the 2D-KS method uses a non-parametric test as the 2D-KS methods does not require a choice of number of bins or the bin size comparing to the Voronoi-Binning method. More importantly, the 2D-KS methods uses the cumulative distribution of stars on the CMD plane rather than assuming independence between sub regions on the CMD plane. This makes the 2D-KS method more robust against the photometric uncertainty caused by the PSF models which cannot be quantified. 

\subsection{Bootstrap Resampling}
\citet{babu_astrostatistics_2006} suggests that for a multivariate KS test, the distribution of KS statistics varies with the underlying true distribution. As a result, the 2D-KS statistics studies by several astronomical literature \cite[e.g.][]{peacock_two-dimensional_1983, fasano_multidimensional_1987} cannot be applied directly to our case. Instead, we combine the method to determine the empirical $\chi^2$ distribution \citep{ying_absolute_2023} with the parametric bootstraping method \citep{babu_astrostatistics_2006}. The bootstrap resampling for the 2D-KS method can be formalized as: let $\left\{ F (.:\theta) : \theta \in \Theta \right\}$ be a family of continuous distributions of both stellar evolution parameters $\theta_{\textup{DSEP}}$, distance $\theta_{\textup{DM}}$, and reddening $\theta_{\textup{RED}}$. The observed data on the CMD plane $X_1, ..., X_n$ comes from the ECDF $F = F(.; \theta)$ for some $\theta = \theta_0$. We calculate the testing statistics $L = \sup_x\left\vert F(x) - F(x; \theta ) \right\vert$ for every set of $\theta$ and assign the estimated parameter $\hat{\theta} = \theta$ which returns the lowest $L$. 

\begin{figure}
    \centering
    \includegraphics[width=0.45\textwidth]{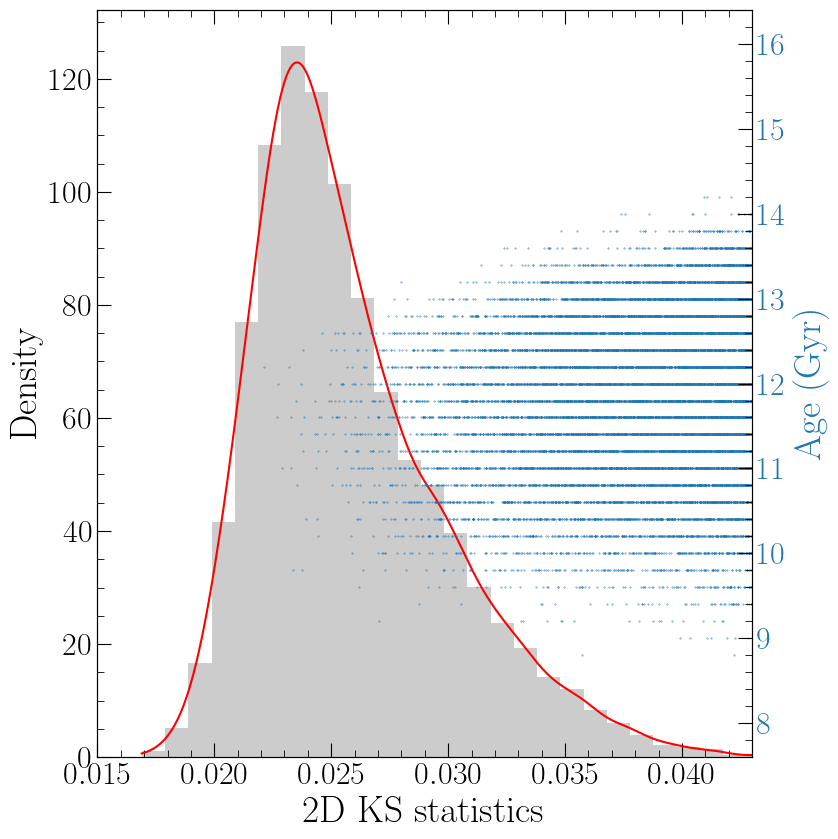}
    \caption{Bootstrap resampling for Voronoi-Binning method. The distribution of the 2D-KS statistics from $10,000$ bootstrap resampling is shown in the grey histogram with estimated kernel density function in red. Blue dots shows the distribution of $\chi^2$ when fitting theoretical isochrones with observed data.}
    \label{fig:2dks_resample}
\end{figure}

A set of simulated observational data $X^*_1, ..., X^*_n$ are generated on the sCMD plane using $\hat{\theta}$. Given the simulated observed ECDF and the underlying distribution parameterized by $\hat{\theta}$, we can generate sCMD using $\hat{\theta}$ and calculate the expected ECDF: $F^*(.;\hat{\theta})$. The testing statistics is calculated as $L^* = \sup_{x}\left\vert F^*_n (x) - \hat{F}(x)\right\vert$. We construct $10,000$ resamples based on the parametric model for each globular cluster. Figure. \ref{fig:2dks_resample} shows the 2D-KS statistics for NGC 3201 (in grey) which serves as the reference probability distribution of $L$ given $\hat{\theta}$, kernal density estimation for the distribution (in red), and testing statistics (at blue). For each isochrone $\theta^*$, the testing statistics $L^*$ is calculated and the probability $p(L = L^* \vert \hat{\theta})$ can be found in the reference distribution. 
\section{Results} \label{Results}
\subsection{Age Estimation}

\begin{figure}[]
    \centering
    \includegraphics[width=0.45\textwidth]{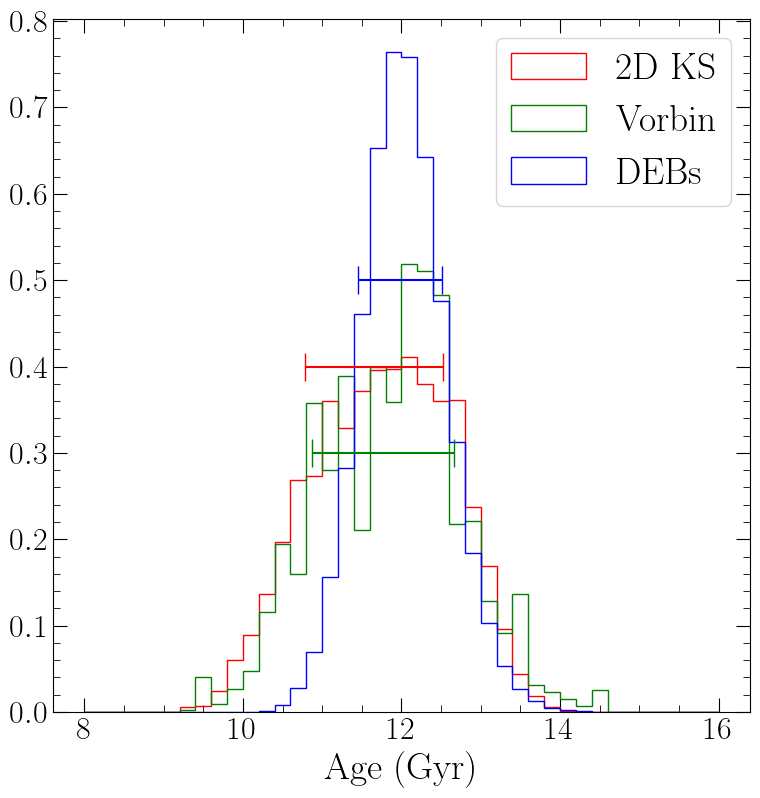}
    \caption{The weighted distribution of ages corresponding to best-fit isochrones based on DEBs (in blue), Voronoi-Binning method (in green) and 2D Kolmogorov–Smirnov method (in red).}
    \label{fig:Age_estimation}
\end{figure}

\begin{table*}[]
\centering
\caption{Summary of results \label{tab3}}
\begin{tabular}{llll}
\hline
Source  & Age (Gyr)    & Distance  Modulus  & Reddening  \\
\hline
DEBs & $11.98 \pm 0.53$& NA & NA\\
Vorbin & $11.76 \pm 0.89$& $14.12 \pm 0.07$& $0.23 \pm 0.02$\\
2D-KS & $11.65 \pm 0.87$& $14.13 \pm 0.08$&$0.24 \pm 0.02$\\
\hline
Combined & $11.85 \pm 0.74$& $14.13 \pm 0.07$& $0.24 \pm 0.02$ \\
\hline
\cite{harris_catalog_1996} & NA & $14.20$ & $0.24$ \\
\cite{rozyczka_cluster_2022} & $11.5 \pm 0.5$ & $14.12^{+0.03}_{-0.05}$ & $0.264 \pm 0.002$ \\
\cite{paust_acs_2010} &$12.0$ & $14.20$ & $0.30$ \\
\cite{valcin_inferring_2020} & $13.05^{+1.05}_{-1.19}$ & $14.20$ & $0.24$ \\
\cite{monty_gemsgsaoi_2018} & $12.2 \pm 0.5$ & $14.27 \pm 0.09$ & $0.25 \pm 0.02$\\
\cite{bono_new_2010} & $11.5 \pm 1.99$  & $14.10 \pm 0.11$& $0.24 \pm 0.02$ \\
\hline
\end{tabular}
\end{table*}

To estimate the absolute age of NGC 3201, we assign weight of each isochrone based on the probability of their corresponding $\chi^2$ in empirical distribution generated from bootstrap resampling. The weight represent the possibility of getting the corresponding $\chi^2$ if our theoretical isochrone represents the underlying population for the observed data.

Figure~\ref{fig:Age_estimation} shows the weighted age distribution from all three methods we used in this study. Figure~\ref{fig:Age_estimation} demonstrate the consistency of age estimated based on $2$ types of observational data using $3$ different fitting methods. Table~\ref{tab3} summarizes the main result for this paper. We compare our result with literature values and our results agrees with previous studies. It it worth noticing that most of the previous CMD based analysis do not take into consideration of uncertainty introduced by stellar evolution models or photometry and usually assume a fixed distance modulus and reddening. We estimate the absolute age for NGC 3201 and take into consideration of uncertainties introduced by those factors and our CMD-fitting methods are able to utilize the number density of stars on CMD to make an estimation with high precision that has uncertainty level comparable with those studies.

\subsection{Monte Carlo Parameters}

\begin{figure}[]
    \centering
    \includegraphics[width=0.45\textwidth]{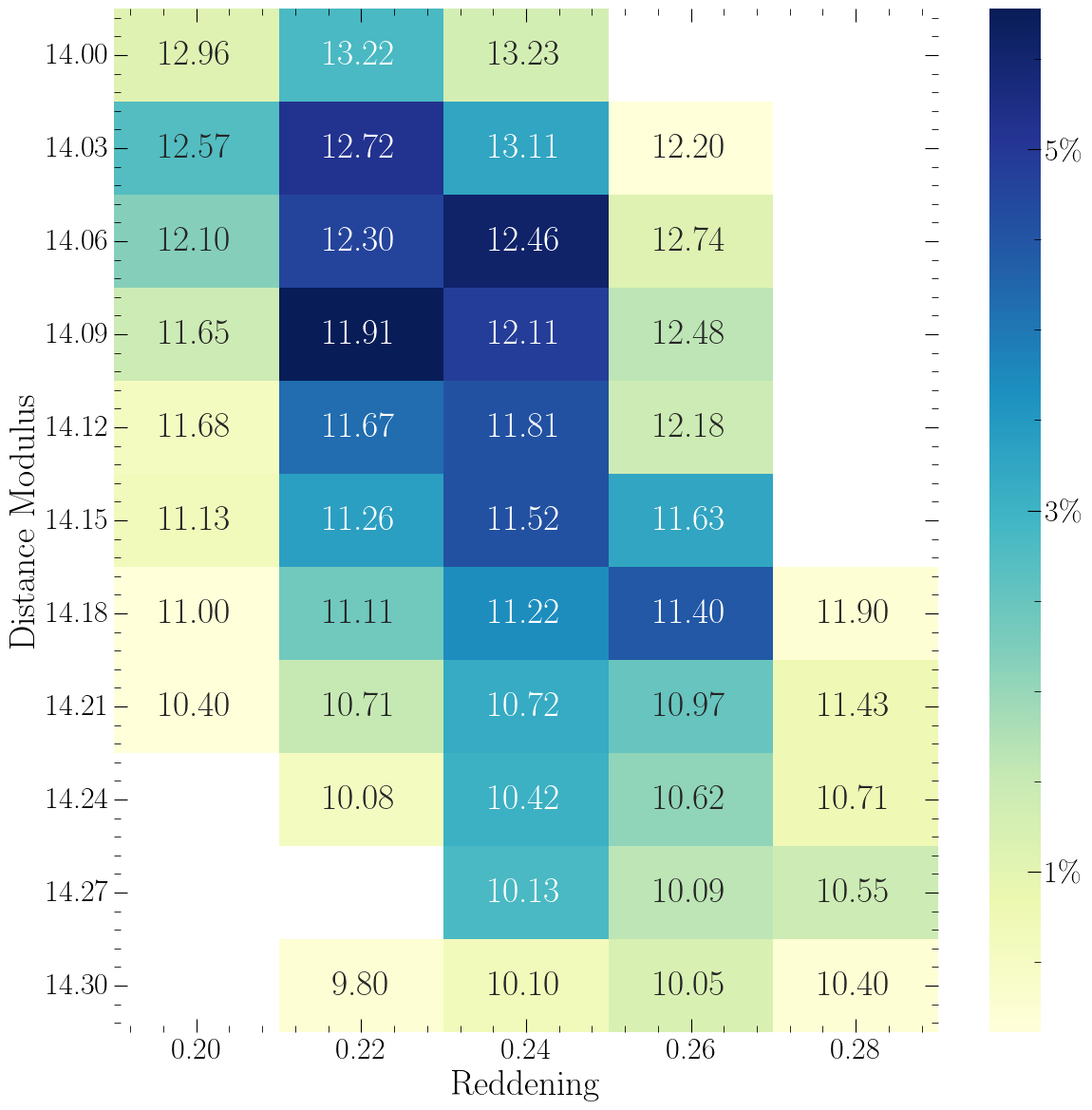}
    \caption{The best-fit age correspond to each combination of distance modulus and reddening. The annotation on each block represents estimated age (Gyr) from each combination. The color represents the occurrence of each combination as a percentage of the best-fit isochrones. }
    \label{fig:Vorbin_DM}
\end{figure}

As described in section \ref{Voronoi-Binning Method}, we test distance moduli ranging from $14.00$ to $14.30$ and reddening ranging from $0.20$ to $0.30$ for each isochrone. The best fitting age corresponding to each distance modulus and reddening bin is shown in Figure \ref{fig:Vorbin_DM}. There is a strong negative correlation between distance modulus and estimated age which is well known.

\begin{figure}[]
    \centering
    \includegraphics[width=0.45\textwidth]{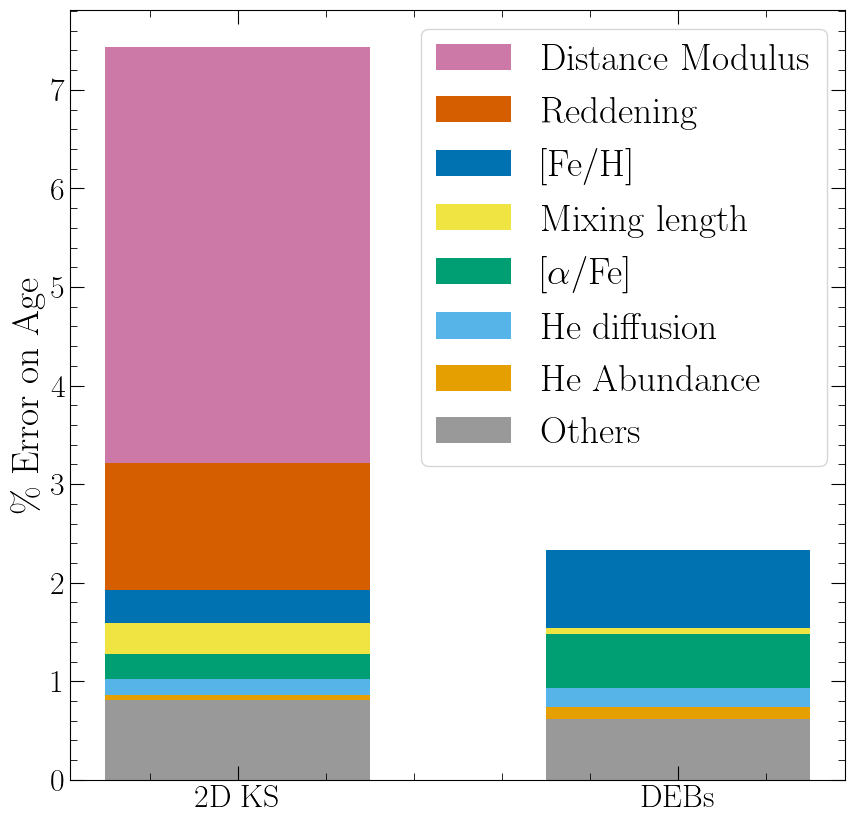}
    \caption{Comparing contributions to the variability of the estimated absolute age of NGC 3201 from each of the Monte Carlo stellar evolution parameters as well as distance modulus and reddening. $\%$ of error on age is determine by the Johnson indices multiplied by the coefficient of variance of the absolute age determined by either method.} We find the variable-importance measure for the 2D KS test results is compatible with that for the Voronoi-Binning method. This is expected as both methods are CMD-fitting methods and are based on the same set of observational data.
    \label{fig:error_budget_johnson}
\end{figure}

It is important to understand each MC parameter's contribution to the variability of the absolute age. Adding finer constraints on important MC parameters, those contribute more to the variability of the response, can produce more precise estimation of the absolute age. Traditional error propagation methods in analyzing contribution to the variability rely on closed form model specification, which is lacking due to the complex nature of stellar evolution. Thus we perform a first-order analysis with a linear regression model specification with the absolute age and the MC parameters. A variety of methods can then be used to decompose $R^2$, the coefficient of determination, by exhibiting hierarchy between the inputs among predictors regarding some dominance criteria known as the general dominance analysis \citep{clouvel_review_2023}.

\citet{lindeman_introduction_1980} proposed a dominance analysis method based on the measure of the elementary contribution of any
given variable  $X_j$ to a given subset model $Y(X_u)$ by the increase in $R^2$ that results from adding that predictive variable to the regression model:
\[
\textup{LMG}_j = \frac{1}{d!} \sum_{\substack{\pi \in \textup{permutation} \\ \textup{of} \{1, ..., d\}}} r^2_{Y,(X_j\vert X_{\pi})},
\]
where $u$ is a subset of all indices ${1, ..., d}$ and $X_u$ represents a subset of input. The computational complexity grows exponentially as the dimension of input parameters increase. In this case, we have over $20$ input parameters which makes it computationally impossible. \cite{johnson_heuristic_2000} introduces the approximation method utilizing relative weight measures to transform the correlated inputs into uncorrelated variables which significantly reduces the computational cost and is used in this analysis.

Figure.~\ref{fig:error_budget_johnson} shows the result in which the uncertainty in the absolute age of both 2D KS method (with coefficient of variance $= 7.44\%$) and DEBs (with coefficient of variance $= 2.33\%$) being decomposed according to the contribution from different parameters. Because the choice of atmosphere model cannot be represented in the linear regressor, we does not include it in the analysis. \citet{ying_absolute_2023} find that the mixing length and atmosphere models are highly correlated and we use the result of mixing length to infer the contribution of uncertainty from the choice of atmosphere models. Figure.~\ref{fig:error_budget_johnson} suggests that distance is the dominant source of uncertainty in Hess fitting which agrees with \citet{ying_absolute_2023}. Because DEB analysis does not assume a distance, and the reddening is already corrected for the observed data, it has a much smaller combined uncertainty (see Table.\ref{tab3}). We found that [Fe/H], Helium abundance, [$\alpha$/Fe], mixing length, and He diffusion are the stellar evolution parameters which are the main source of uncertainties for both Hess fitting and DEBs analysis which further demonstrate the consistency of our methods. It is worth notice that there is a significant more contribution from Helium abundance for DEBs analysis than for Hess fitting. Helium abundance is important to stellar evolution as a helium rich star evolves faster, and at a higher temperature and luminosity. Despite using different methods on different data sets, there is a consistent $\sim 0.6\%$ of uncertainty in age which cannot be explained by any MC parameters which might be the linear regression model not being able to describe the complex parameter structure or the parameter misspecification in our 1D stellar evolution model. 

\section{Conclusion}
We determine the absolute age of NGC 3201 with $2$ independent sets of observational data and $3$ different statistical analysis methods. We apply a Monte Carlo simulation approach to take into consideration of uncertainty introduced by stellar evolution parameter, distance modulus and reddening. We create $10,000$ sets of MC generated stellar evolution models  with $21$ variables  using Dartmouth Stellar Evolution Program \citep{dotter_dartmouth_2008}, an state-of-art 1D stellar evolution code. We construct theoretical isochrones from $8$ Gyr to $16$ Gyr with $0.2$ Gyr increment for each set of input parameters based on the framework described by \citet{dotter_mesa_2016}. 

Every isochrone constructed is used to fit the DEBs in NGC 3201 with a $\chi^2$ based goodness-of-fit testing method. The results are compared with the distribution of the same metric in a bootstrap resampling, and being converted to a probability. 

Each isochrone constructed is also used to generate a sCMD with $4,000,000$ data points. Two different statistical methods are applied to compare the sCMD with observed CMD created from HST ACS data. The Voronoi-Binning method divide the CMD into $200$ sub-regions and compare the number density of observed and simulated data while the 2D-KS test method utilize the ECDF to detect the biggest change in morphology between observed and simulated data. In both methods, we introduce shift in distance modulus ranging from $14.00$ to $14.30$, and reddening ranged from $0.20$ to $0.30$ and use a Gaussian Process approach to get the combination of distance and reddening which lead to the best fit. 

The  absolute age of NGC 3201 from $3$ different statistical analysis methods with $2$ independent sets of observational data agree with each other. The results are combined and the absolute age of NGC 3201 $= 11.82 \pm 0.66$ Gyr. We perform variable importance analysis on both Hess fitting and DEBs analysis results. We found that distance is the most dominant source of uncertainty for the absolute age of NGC 3201 in Hess fitting which is in line with our conclusion for analysis of M92 \citep{ying_absolute_2023}. Because DEBs analysis does not rely on an assumption of distance, the uncertainty of DEBs analysis is $~50\%$ of the uncertainty of CMD-fitting despite having a much smaller sample size. Moreover, we find the metallicity, $\alpha$ enhancement, mixing length, and treatment of helium diffusion being the most important stellar evolution parameters for both DEBs analysis and CMD-fitting. 

The absolute age of NGC 3201 with $\mathrm{[Fe/H]} = -1.5$ is found to be $2.0\pm 1.0\,$Gyr younger than M92 with $\mathrm{[Fe/H]} = -2.3$. This is suggestive, though not conclusive evidence that the comparatively metal-rich cluster NGC 3201, which was likely accreted by the Milky Way, formed at a significantly lower redshift than the metal-poor Milky Way cluster M92.  Future work, using a larger sample of GCs with a range of metallicities, can test this tentative conclusion.

\section*{Data Availability}
All the mission data used in this paper can be found in MAST: \dataset[https://doi.org/10.17909/t9tg65]{https://doi.org/10.17909/t9tg65}.
The $10,000$ sets of Monte Carlo isochrones constructed for this study along with the differential reddening corrected photometric data for NGC 3201 are available on Zenodo under an open-source Creative Commons Attribution license: \dataset[doi:10.5281/zenodo.10729299]{https://doi.org/10.5281/zenodo.10729299}

\section*{Acknowledgments}
This material is based upon work supported by the National Science Foundation under Award No.~2007174, by NASA through AR 17043 from the Space Telescope Science Institute (STScI), which is operated by AURA, Inc., under NASA contract NAS5-26555.

\bibliographystyle{aasjournal}
\bibliography{NGC3201_reference_final}
\end{document}